# Network Medicine in the age of biomedical big data

**Abhijeet R. Sonawane[1,2], Scott T. Weiss [1,2], Kimberly Glass[1,2], Amitabh Sharma[1,2,3]***

[1]Channing Division of Network Medicine, Brigham and Women's Hospital, Boston, MA 02115, USA

[2]Department of Medicine, Harvard Medical School, Boston, MA, USA

[3]Center for Interdisciplinary Cardiovascular Sciences, Cardiovascular Division, Brigham and Women's Hospital, Boston, MA, USA

**\* Correspondence:**
Corresponding Author
amitabh.sharma@channing.harvard.edu

## Abstract

Network medicine is an emerging area of research dealing with molecular and genetic interactions, network biomarkers of disease, and therapeutic target discovery. Large-scale biomedical data generation offers a unique opportunity to assess the effect and impact of cellular heterogeneity and environmental perturbations on the observed phenotype. Marrying the two, network medicine with biomedical data provides a framework to build meaningful models and extract impactful results at a network level. In this review, we survey existing network types and biomedical data sources. More importantly, we delve into ways in which the network medicine approach, aided by phenotype-specific biomedical data, can be gainfully applied. We provide three paradigms, mainly dealing with three major biological network archetypes: protein-protein interaction, expression-based, and gene regulatory networks. For each of these paradigms, we discuss a broad overview of philosophies under which various network methods work. We also provide a few examples in each paradigm as a test case of its successful application. Finally, we delineate several opportunities and challenges in the field of network medicine. Taken together, the understanding gained from combining biomedical data with networks can be useful for characterizing disease etiologies and identifying therapeutic targets, which, in turn, will lead to better preventive medicine with translational impacts on personalized healthcare.

# 1 Introduction

Biological systems are comprised of various molecular entities such as genes, proteins and other biological molecules, as well as interactions between those components. Understanding a given phenotype, the functioning of a cell or tissue, etiology of disease, or cellular organization, requires accurate measurements of the abundance profiles of these molecular entities in the form of biomedical data. Analysis of the biomedical data allows us to explain important features of the interactions leading to a mechanistic understanding of the observed phenotype. The interplay between different components at different levels can be represented in the form of biological networks, for example, protein-protein interactions (PPI) (Cusick et al., 2005; Uetz et al., 2000) and gene regulatory networks (Davidson, 2006). Different biological networks capture the complex interactions between genes, proteins, RNA molecules, metabolites and genetic variants in the cells of organisms. These networks, also interchangeably known as graphs, are representations in which the complex system components are simplified as nodes that are connected by links (edges) (Vidal et al., 2011). Networks provide a conceptual and intuitive framework to model different components of multiple omics data from the genome, transcriptome, proteome, and metabolome (Figure 1) (Liu and Lauffenburger, 2009).

The convenient representation of the biological components in graphs led to the field of network biology - a discipline that studies holistic relationships between various biological components by combining graph theory, systems biology, and statistical analyses (Lindfors, 2011; Walhout et al., 2012). Moreover, the quantitative tools of network biology offer the potential to understand cellular organization and capture the impact of perturbations on these complex intracellular networks (Wang et al., 2011). Network Medicine is an extension of network biology with a set of focused goals related to disease biology, including understanding disease etiology, identifying potential biomarkers, and designing therapeutic interventions, including drug targets, dosage, and synergism discovery (Loscalzo et al., 2017). Research in network medicine heavily depends on large datasets for building models, making predictions and assessing their validity. The promise of network medicine research is to develop a more global understanding of how perturbations propagate in the system by identifying the pathways, sub-types of disease states, and key components in the networks that can be targeted in clinical interventions. Moreover, networks are the centerpiece of the 'new biology' in the biomedical data revolution and translation to personalized medicine (Schadt and Bjorkegren, 2012).

Advances in high-throughput biotechnologies have led to the generation of massive amounts of biomedical data that provides new research avenues. The rapid decline in costs due to technological advancements such as next-generation sequencing (NGS) have provided the necessary impetus to generate multiple large-scale multi-omics biomedical data-sets that characterize various phenotypes. This includes exome and whole genome sequencing, transcriptomics, proteomics, lipidomics, microbiomics etc. (Schadt and Bjorkegren, 2012). Constructing appropriate network models is a challenging problem that heavily depends on the study design, the phenotype under study, the molecular entities measured, and the type and size of the data. The field of network medicine is largely discovery — rather than hypothesis — driven, uncovering previously unknown relationships and leading to the identification of new biomarkers. The statistical rigor of network predictions comes from the study design and the size of the datasets. Large-scale consortium-based efforts looking at the various aspects of human biology have allowed the application of network-based methods to uncover new insights into the molecular mechanisms of the given phenotype, such as tissue specificity or disease context. In this review, we first examine various large-scale biomedical datasets and types of biological networks as summarized by Figure 1. We then provide three paradigms in which biological networks can be combined with big biomedical data to understand the given phenotype.

## 2      Biomedical data sources

Recent technological advancements in sequencing technologies, resulting in a reduction in cost per base pair, have heralded an era of massive data generation for different types of molecular profiles across a broad range of phenotypes and diseases. After the completion of the human genome project (Collins et al., 2003), HapMap project (International HapMap, 2003) created an extensive catalogue of common human genetic variants, the differences in DNA sequences, based on microarray data. These studies eventually progressed into the '1000 Genomes Project' (Genomes Project et al., 2015), which leveraged NGS technologies. In cancer research, The Cancer Genome Atlas (TCGA) (Cancer Genome Atlas Research, 2008) contains profiles of tumors and matched normal samples from more than 11000 subjects for 33 cancer types. The repertoire of TCGA data includes clinical information (demographic, treatment, and survival information), gene expression profiling, microRNA profiling, copy number variation (CNV) (genomic structural variations) identifications, single nucleotide polymorphism (SNP), DNA methylation (whole genome methylation calls for each CpG site), and exon sequencing (expression signal of particular composite exon of a gene). Together these data have helped in the identification of driver somatic mutations, the molecular basis of cancer progression, and potential therapeutic interventions for cancer subtypes. To understand the role of the epigenetic state in gene regulation and to characterize the functional elements of the transcriptional machinery, the Encyclopedia of DNA Elements (ENCODE) consortium for humans (Consortium, 2012), model organism Encyclopedia of DNA Elements (modENCODE) for model organisms (Yue et al., 2014), and ROADMAP Epigenomics project (Romanoski et al., 2015) were commissioned to improve the understanding of how epigenomics contributes to disease. The Riken-led Functional ANnoTation Of Mammalian genome (FANTOM5) (Andersson et al., 2014) project provided cell-type-specific enhancer elements and identified pathobiological regulatory SNPs. To further understand transcriptional patterns in human tissues and their relationship with the genotype, Genotype-Tissue Expression (GTEx) data was generated (Consortium, 2015; Mele et al., 2015). Trans-Omics for Precision Medicine (TOPMed) (Prokopenko et al., 2018) is another set of multi-omics data on 100k individuals that also includes clinical data and is aimed at understanding the fundamental biological processes that underlie heart, lung, blood, and sleep disorders. The Precision Medicine Initiative or "All of Us" program (https://allofus.nih.gov/) aims to acquire a broad range of data from about 1 million individuals.
Since 2003, the Human Protein Atlas (HPA) (Uhlen et al., 2005; Uhlen et al., 2015) by Swedish consortium has been releasing data on protein expression levels in cells, tissues, and various pathologies, including 17 cancer types. Similarly, the Human Cell Atlas (HCA) (Rozenblatt-Rosen et al., 2017) aims to provide a reference map of single cell omics data in human cells and cell types. The UK-Biobank (Allen et al., 2014; Sudlow et al., 2015) is another commercial resource that has an array of health-related measurements on patients, including biomarkers, images, clinical information, and genetic data. The Human Microbiome Project (HMP) (Turnbaugh et al., 2007) is a categorization of microbiota on different human body sites whose goal is to understand the role of the microbiome and the impact of its dysbiosis on human disease. Apart from these large international databases looking at one or more aspects of health or disease, many other resources from the concerted efforts over decades of data collection are also available. This includes the Nurses' Health Study (Belanger et al., 1978; Colditz et al., 2016), Health Professionals Follow-up Study (Grobbee et al., 1990), Framingham Heart Study (Dawber et al., 1951; Mahmood et al., 2014), and COPDGene (Pillai et al., 2009). This wealth of biomedical data not only allows for a deeper probing of the underlying biological systems, but also inspires the development of novel methods that can maximize the information that can be extracted from these data. The tools developed within the field of network medicine are highly versatile, enabling their customized application depending on the given biological or disease context.

Collecting large-scale multi-time point data across multiple omics in different disease conditions is expensive and often not feasible, especially for human subjects. However, small-scale longitudinal for a single omic, such as gene expression, is available in biomedical databases (Bouquet et al., 2016; Jung et al., 2015). High resolution mass spectrometry has also allowed for the collection of longitudinal proteome data, for example to test the effect of drugs (Fournier et al., 2010) or oxidative stress (Vogel et al., 2011) in yeast. A longitudinal multi-omic dataset containing both human transcriptomic and proteomic information has been analyzed to study changes in molecular profiles (Chen et al., 2012). Multi-omic datasets such as this one allows us to probe the relationship between biological molecules based on the central dogma of biology, such as the connection between transcript abundance and protein levels (Liu et al., 2016; Marguerat et al., 2012). Longitudinal data is also amenable to temporal or dynamical network analysis, wherein one can evaluate the statistical dependence of the state of a network on the gene expression patterns from previous time steps (Dondelinger and Mukherjee, 2019; Kim and Kim, 2018). Kim et al. provide a summary of several methods to infer temporal regulatory relationships (Kim et al., 2014).

In the next section, we will review some of the main types of biological networks constructed using high throughput molecular profiling, literature mining, or manual curation of scientific literature.

## 3    Primer on biological networks

Each network-based study has to primarily identify two things: what are the critical entities in the system under investigation (nodes), and what is the nature of the interactions between these entities (edges) (de Silva and Stumpf, 2005). This information often comes from multiple different data sources, dealing with the various facets of the biological system. For example, Protein-Protein Interactions (PPIs), also defined as the interactome, is a network of proteins and physical interactions between them (Cusick et al., 2005). These interactions can be obtained from yeast-2-hybrid assays (Li et al., 2004; Vidal and Fields, 2014), co-immunoprecipitation (Lin and Lai, 2017), literature text-mining (Papanikolaou et al., 2015), 3D structure (Lu et al., 2013), co-expression of genes (Bhardwaj and Lu, 2005), sequence homology (Shen et al., 2007), and other sources. Each of these data sources have both merits and demerits (Cusick et al., 2005). These networks inform us about the overall topological properties of protein interactions as well as the positions of specific proteins within this network. However, extracting phenotype specific (i.e. cell, tissue or disease-specific) information based on the PPI remains an open challenge and requires the development of novel ways of integrating biomedical data with these networks.

Gene co-expression networks (GCNs) and gene regulatory networks (GRNs) often make direct use of phenotype-specific gene expression data in the network construction, with additional analysis required to extract meaningful biological information for the underlying phenotype. The availability of transcriptomic data for a wide range of phenotypes presents an opportunity to probe the patterns of molecular co-abundance, albeit with limitations concerning the interpretation of the biology. GCNs can be constructed in many ways, including information theoretic, regression-based, and Bayesian approaches (Butte and Kohane, 1999). Several common methods for constructing GCNs include Weighted Gene Co-expression Network

Analysis (WGCNA) (Langfelder and Horvath, 2008), Context Likelihood of Relatedness (CLR) (Faith et al., 2007), Algorithm for the Reconstruction of Accurate Cellular Networks (ARACNe) (Margolin et al., 2006), Partial Correlation and Information Theory (PCIT) (Reverter and Chan, 2008), Gene Network Inference with Ensemble of Trees (GENIE3) (Huynh-Thu et al., 2010), Supervised Inference of Regulatory Networks (SIRENE) (Mordelet and Vert, 2008), and Gene CO-expression Network method (GeCON) (Roy et al., 2014). Gene regulatory networks (GRN) are a related type of network that attempts to look beyond the co-abundance of gene expression and instead identify the influencing patterns of transcription factor genes over others in a mechanistic fashion (Marbach et al., 2012). Since transcriptional regulation depends on cis and trans-regulatory elements as well as transcription factor binding, GRNs often incorporate this information during model construction. Many methods with a modified definition of correlations have been proposed to infer GRNs. However, identifying the putative cis-regulatory sequences, such as those found in the promoter regions of genes, that are relevant for a specific biological context is important to enable the understanding of disease, tissue, or cell-specific regulatory perturbations. The location of TF binding to the DNA can be assayed using yeast-1-hybrid (Deplancke et al., 2004), ChIP-Seq (Jaini et al., 2014), or inferred by other means (Mundade et al., 2014). However, the cost and other limitations involved in generating these data in a context-specific manner have meant that incorporating this information when constructing putative regulatory networks remains a challenge.

Other types of biological networks include metabolic networks, which represent a collection of biochemical interactions between metabolites and enzymes (Terzer et al., 2009). Ecological networks, which represent biotic interactions, can also be applied to microbiome data, the collection of microbes' genes, to construct microbiome networks (Bauer and Thiele, 2018; Coyte et al., 2015; Layeghifard et al., 2017; Rottjers and Faust, 2018). Together, genotype and transcriptomic data can be used to map genetic variants to genes and then summarized in an expression Quantitative Trait Loci (eQTL) network (Fagny et al., 2017; Platig et al., 2016). A network of immune cell communication has been constructed using high-resolution mass spectrometry-based proteomics data and was shown to exhibit social networks-like properties. Disease networks, also known as the diseasome, have been proposed; these networks connect diseases and disorders with disease genes based on Online Mendelian Inheritance in Man (OMIM) associations (Boyadjiev and Jabs, 2000; Goh et al., 2007; Hamosh et al., 2002; Wysocki and Ritter, 2011; Zhang et al., 2011). Similarly, networks connecting symptoms with diseases have helped to shed light on the shared genetic associations between diseases (Zhou et al., 2014). Efforts to identify specific disease-causing genes, using genomic intervals obtained from linkage mappings or Genome-Wide Association Studies (GWAS), have been undertaken using hybrid heterogeneous networks. These hybrid networks often include a combination of disease-gene networks, generic or tissue-specific molecular networks such as PPIs or GCNs, and prior knowledge of disease similarities (Moreau and Tranchevent, 2012; Navlakha and Kingsford, 2010; Ni et al., 2016). Various network-based tools have been implemented in the gene prioritization problem (Li and Patra, 2010; Tian et al., 2017; Wu et al., 2008). All these aforementioned types of network biology approaches are particularly useful in understanding complex diseases, which result from multiple genetic factors and environmental influences (Moreau and Tranchevent, 2012).

Analysis of biological networks also necessitates understanding their structural or topological properties. This includes the identification of important modulators, driver nodes, local network structures, and recurrent subgraphs in the network. Local connectivity properties such as degree and other centrality metrics can help to identify key molecular entities that dominate various network neighborhoods, such as hubs, bottlenecks, or core nodes. At the global level, properties like average path length, degree distribution, diameter, clustering coefficients, and controllability (Liu et al., 2011) help with the characterization and comparison of network topologies. Mesoscale measures such subgraphs or network motifs – recurrent patterns connecting a fixed number of nodes (typically 3 or 4) – are considered fundamental components of biological networks (Milo et al., 2002). An extension of network motifs to include more nodes, or graphlets, has been used to analyze the interactome (Davis et al., 2015; Malod-Dognin et al., 2017; Przulj et al., 2004). Identifying the connectivity patterns enriched in a network (i.e. over-represented with respect to a null model) can help to compare, characterize, and discriminate between networks (Alon, 2007; Przulj, 2007; Shen-Orr et al., 2002). These patterns are also commonly associated with control substructures that dominate information flow in the networks, especially in transcriptional regulatory, neuronal, and social networks.

## 4    Integrating data with networks/Combining biomedical data with networks: challenges and ways

The ultimate aim of inferring biological networks using biomedical data is to provide lab-testable hypotheses by identifying biomolecular entities that play a crucial role in the observed phenotype (Figure 1). Detecting changes in abundance levels of these biomolecules and their interaction landscape in the context of a tissue, cell, or disease-specific environment requires both relevant data and the application of appropriate network analysis. Each biological network analysis has strengths and limitations based on how it incorporates phenotype specific data, and the research question being addressed (Altaf-Ul-Amin et al., 2014; Kanaya et al., 2014). In some cases, it is possible to identify a baseline network from general physical interactions between proteins, after which disease or phenotype-specific information from specific experiments can be overlaid to generate a more context-specific network.

PPI networks provide a fabric of potential interactions between proteins, but phenotype-specific interactions can only be added as an extra layer from separate biomedical data. The hypothesis behind analyzing such networks, combination of baseline PPI with disease information added as next step, is that the defects or mutations in only a few genes or proteins may propagate to other components in the network, and that this collection of affected genes constitute a critical module in the network (Schadt and Bjorkegren, 2012). Previous work along these lines has shown that these modules are not only structurally related but are also functionally relevant to the observed phenotype. This central tenet of network medicine from the interactome has been successfully tested for many diseases and other phenotypes (Goh et al., 2007; Huang et al., 2018b; Huttlin et al., 2017; Lim et al., 2006; Menche et al., 2015; Sahni et al., 2015; Sharma et al., 2013; Sharma et al., 2018a; Sharma et al., 2015; Taylor et al., 2009; Wang et al., 2018; Willsey et al., 2018) and has also led to novel drug-target discoveries (Guney et al., 2016; Luo et al., 2017; Yildirim et al., 2007) along with novel interactions between genes. Despite recent advances, the PPI is incomplete and inferring disease-specific interactions requires innovative strategies in order to overcome this discrepancy.

GCNs are by definition context-specific, as they are constructed by calculating correlations in a given gene expression data set. In contrast, GRNs often are built starting from a baseline network composed of all potential interactions between transcription factors and genes. This baseline network can be derived from genetic sequence information and DNA-binding domain sequences within regulatory proteins, such that an interaction is inferred if a given gene's promoter contains the binding motif of a particular TF. Disease or tissue-specific information then has to be integrated with this baseline prior network to obtain meaningful information about perturbations caused due to the disease.

In this review, we explore PPI, GCNs, and GRNs, and also provide exemplar methods for each. Based on these three types of networks, we describe three complimentary philosophies and

*modus-operandi* to embed phenotypic specific molecular information from biomedical data into a network framework, as shown in Figure 2. We present these paradigms to demonstrate that applying network phenomenology to big biomedical data requires a nuanced, condition-specific approach. In the following sections, we will focus on each paradigm separately, providing their examples, the questions they intend to answer, and the diagnostics of the outcomes. We mainly focus on reviewing the methods to integrate multi-omic data to extract phenotype specific information, specifically the disease and tissue specificity in PPI, GCN, and GRN.

## 4.1 PARADIGM I: Network-based approach to human disease using the interactome

The high-throughput mapping of the interactome has provided a molecular interaction map of the genes encoding proteins that might drive an underlying pathophenotype (Barabasi et al., 2011; Hein et al., 2015; Huttlin et al., 2015; Kamburov et al., 2009; Rolland et al., 2014; Zhang et al., 2013). Understanding disease associated biomedical data in the context of the network principles supports the discovery of more accurate biomarkers, localization of the disease perturbation in the network, personalized networks, better disease sub-type classifications, better targets for drug developments, and better drug repurposing. Using this paradigm, one can extract disease-specific signals in a variety of ways. One may consider topological properties of the nodes and assess the functional role of their *hubness*, i.e., a node property of having higher number of connections. Alternatively, one can also identify new disease genes in the network by using 'guilt-by-association' (Aravind, 2000; Huang et al., 2018b; Lage et al., 2007; Lee et al., 2011; Quackenbush, 2003; Sharma et al., 2010; Sharma et al., 2013; Stuart et al., 2003) — a property ascribed not based on direct evidence but association with other disease genes, albeit with care (Gillis and Pavlidis, 2012). In addition to prioritizing candidate disease genes, molecular interaction networks can assist in identifying the sub-networks that are mechanistically linked to disease phenotypes (Emamjomeh et al., 2017; Menche et al., 2015; Sharma et al., 2015; van Dam et al., 2018). The proteins in these connected subnetworks may have clinical importance by being therapeutic targets and biomarkers (Sharma et al., 2015). Network tools can also provide a framework for disease classification (Halu et al., 2017; Zhou et al., 2018).

Assessing disease genes from other, non-disease genes by their topological properties on the interactome have provided new insight into disease pathobiology. It was found that disease genes tend to have non-hub properties (Goh et al., 2007). Later, it was reported that genes from OMIM and those associated with cancer are more central in a literature-curated interactome (Ideker and Sharan, 2008; Jonsson and Bates, 2006; Xu and Li, 2006). Further, several studies demonstrated that disease genes, in general, mostly have a high-degree and a low clustering coefficient (number of mutual connections with the neighboring nodes) (Cai et al., 2010; Feldman et al., 2008). Moreover, recently it was reported that disease genes have a higher degree, but it was discovered that the cancer-related genes are the primary drivers of this trend (Jonsson and Bates, 2006; Wachi et al., 2005). Genes associated with either Mendelian or complex diseases also have higher degree and lower clustering coefficients compared to non-disease genes (Cai et al., 2010; Pinero et al., 2016). The topological properties of disease-associated genes vary significantly from disease to disease. The factors that influence these discrepancies include the incompleteness of the current interactome, bias toward well-studied genes, and incomplete knowledge about the number genes associated with various diseases (Menche et al., 2015). It is anticipated that the alliance of different technologies like yeast-2-hybrid, affinity purification mass-spectrometry (AP-MS), and cross-linking AP-MS (Schweppe et al., 2018) will provide access to larger data that will be helpful in providing knowledge about the missing interactions. On the disease-gene discovery side, projects like the UK biobank prospective cohort study, which includes in-depth genetic and phenotypic data, will enhance knowledge regarding the missing disease genes (Bycroft et al., 2018).

An important area in which the interactome has helped in understanding complex diseases is the prediction of disease associated genes. The goal is to identify novel genes and proteins, which are involved in the regulation of tissues, or dysregulated in case of disease, through the association with observed disease candidate genes using the biological hierarchy of molecular interactions. Figure 2A depicts this paradigm where the PPI network serves as map of potential biological interactions between various proteins over which disease associates genes are mapped to uncover relevant biology. The central philosophy in most methods under this paradigm is that the neighbors of the disease associated components or network modules, such as a set of differentially expressed genes (Chuang et al., 2007) or genes with disease-associated SNPs (Barrenas et al., 2012; Feldman et al., 2008; Lage et al., 2007; Oti et al., 2006), could potentially be associated with similar diseases (Goh et al., 2007), and are closer to each other as compared to the other nodes in the network. The definition of this closeness, or vicinity of nodes, just like the definition of modules and clusters, varies with different research strategies. Some methods assume topological closeness in terms of the number of shortest paths connecting given nodes, while others take the similarity of biological function into account. Guilt-by-association methods focus on identifying new disease genes by optimizing based on both the local and global properties of the network and by considering the role of other disease genes and their neighborhood. Network-based strategies to find disease genes and their associated mechanisms can be divided in two types: exploratory and analytic methods (Carter et al., 2013). In exploratory methods one can analyze the biological trends due to perturbations. For example, Chu et al. expanded on known angiogenesis pathways to construct a protein-protein interaction network for angiogenesis (Chu et al., 2012). In contrast, analytic methods aim to identify specific genes and pathways associated with a disease. For example, Gilman et al. developed a method for network-based analysis of genetic associations to identify a biological network of genes affected by rare de nova CNVs in autism (Gilman et al., 2011). Recently, Huang et al. systematically evaluated 21 protein-interactions networks for the ability to recover disease genes sets (Huang et al., 2018b). After correcting for size, they found that the Database for Interacting Proteins (DIP) network (Xenarios et al., 2000) had the highest efficiency in recovering disease genes (Huang et al., 2018b).

In contrast to predicting the disease candidate proteins, finding the associated disease-related network components, or sub-networks, provides a more substantial network space to discover the pathways and mechanisms that influence disease. Goh et al. proposed a correlation between the location of disease-associated genes and the topology of the molecular interaction network (Goh et al., 2007). The tendency of disease-associated genes to interact more often with others compared to random genes in the interactome led to the establishment of the 'local impact' hypothesis (Barabasi et al., 2011). According to this hypothesis, molecular entities involved in similar diseases have an increased tendency to interact with each other and to localize in a specific neighborhood of the interactome (Barabasi et al., 2011). The search for these modules involves exploring the structural and topological properties of the PPI network. Community detection algorithms (Spirin and Mirny, 2003), clique percolation (Sun et al., 2011), and genetic algorithms (Liu et al., 2018) have been applied to uncover disease modules using network properties (Vlaic et al., 2018). Module prediction and identifying non-overlapping clusters with the PPI remains challenging since the PPI network has a short diameter, i.e., most nodes are close to all other nodes in terms of network distance. Novel distance metrics and community detection algorithms have been proposed to overcome this problem (Hall-Swan et al., 2018). The recently proposed DIseAse MOdule Detection (DIAMOnD) algorithm

(Ghiassian et al., 2015) associates the functional modules of known disease-associated proteins (seed proteins) and identifies the close neighbors of these genes (candidate disease-associated proteins) using topological properties of the interactome. The method suggests that the *connectivity significance* among the disease-associated proteins is the best predictive quantity to find the disease related components in the interactome. The underlying hypothesis is that close neighbors of known disease proteins may be involved in the disease. The working principle of DIAMOnD is as follows; first, a pool of disease genes encoding proteins is identified for a disease of interest from biological experiments, genome-wide association studies, linkage analysis, or other disease associated data sources (Pinero et al., 2017). Next, these disease proteins (seeds) are mapped onto the interactome. Next, neighbor proteins are added iteratively to the set of seed proteins based on the condition that each neighbor added is most significantly connected to the seed proteins. A hypergeometric test assigns a p-value to the proteins that share more connections with seed proteins than expected by chance. Finally, the seed proteins plus the added neighbor proteins are part of network components that represent a disease module, or a subnetwork of proteins in the interactome, the members of which are more functionally and topologically related to each other than to other portions of the network. These subnetworks are designated as disease-specific modules based on the source of initial seed proteins. Disease module identification has also led to endophenotypes, intermediate pathophenotypes, and network modules describing their common and distinctive molecular mediators (Ghiassian et al., 2016; Lage et al., 2008).

As mentioned previously, significant progress has been made in mapping the interactome by high-throughput approaches like yeast-2-hybrid (Dreze et al., 2010; Rolland et al., 2014; Rual et al., 2005; Venkatesan et al., 2009), AP/MS (Hein et al., 2015; Huttlin et al., 2017; Huttlin et al., 2015) and various literature-curated data sources, such as ConsensusPathDB, STRING, and PCNet, which collate the known and predicted interactions between proteins (Klingstrom and Plewczynski, 2011). Despite these efforts, the current interactome mapping is 80% incomplete (Hart et al., 2006; Menche et al., 2015; Mosca et al., 2013; Venkatesan et al., 2009) and is affected by many experimental and literature biases. Given the incompleteness of the interactome, it is possible that the disease modules are also far from complete. An attempt to overcome this limitation was made using a network-based closeness approach that compares the weighted distance between different disease and seed-gene neighborhoods to random expectation on the network. In the context of Chronic Obstructive Pulmonary Disease (COPD), 140 potential candidate genes (Sharma et al., 2018) were identified. Another shortcoming of disease module detection related to the lack of context-dependence and tissue-specificity within the PPI was studied by (Kitsak et al., 2016). They found that the genes expressed in a particular tissue tend to form localized connected subnetworks, which overlap between similar tissues and are situated in the different neighborhoods for pathologically distinct pairs of tissues. The perturbations in tissue-dependent subnetworks may help us understand disease manifestations or pathophenotypes. Integrating multi-omics data, including epigenomics, proteomics, and metabolomics, with PPI analysis remains challenging, but is critical for identifying disease or tissue-specific modules in the interactome.

## 4.2 PARADIGM II: Identifying important genes using patterns of co-abundance of biomolecules

Measuring transcript abundance or gene expression patterns for given phenotypes (case-control) across multiple samples is one of the main research strategies used to probe the system as it is connected to the central dogma of molecular biology. Performing differential gene expression analysis often identifies important genes affected by the disease. However, it does not provide information regarding how these genes are influenced by or influence other genes.

It has been observed that genes with similar expression patterns might be part of complexes, influence each other, or be part of the same pathways or mechanisms (Serin et al., 2016). This inspired the construction of gene co-expression networks where the patterns of transcript abundance are studied in the context of the disease. The central philosophy of this paradigm is to combine important seed genes with an organic network of co-expression patterns derived from the gene expression data from the same system.

There are many ways to compute co-expression or co-abundance patterns, including using Pearson correlations (Stuart et al., 2003), Spearman rank correlations (Liesecke et al., 2018; Song et al., 2012), mutual information (Butte and Kohane, 1999; Margolin et al., 2006; Meyer et al., 2007), Gaussian graphical models (Toh and Horimoto, 2002), regression-based methods (Pirgazi and Khanteymoori, 2018; van Someren et al., 2006; Yeung et al., 2002), Bayesian approaches (Friedman et al., 2000; Li et al., 2007; Perrin et al., 2003; Xing et al., 2017), random matrix theory (Jalan et al., 2010; Jalan et al., 2012; Luo et al., 2007), and partial correlations (Reverter and Chan, 2008). Gene co-expression networks (GCNs) identify the functionally coordinated participation of genes in response to an external stimulus or condition. GCNs can be signed or unsigned, weighted or unweighted, and may either be constructed using microarray or RNA-Seq data. Care must be exercised when using thresholding methods to obtain unweighted co-expression networks as these are subjective and can change the network structure and topology (Elo et al., 2007); methods based on the clustering coefficient (Boyadjiev and Jabs, 2000), random matrix theory (Luo et al., 2007), or soft thresholding, which raises the weights by a certain power to penalize weaker edges (Langfelder and Horvath, 2008), have been used to address this limitation. Along with total gene expression levels, isoform abundance and alternative splicing can also be used in constructing gene co-expression networks (Saha et al., 2017).

GCNs are also used to identify co-expression modules. Clusters, modules, or subgraphs of genes that have similar functions are often highly interconnected in GCNs. These clusters can be identified using network topology-based methods like community detection (Girvan and Newman, 2002), modularity maximization (Newman, 2004), K-means clustering (Stuart et al., 2003), or variants of hierarchical clustering methods (Langfelder and Horvath, 2008; Serin et al., 2016). The genes in the most significant modules are then assessed for their biological importance using functional enrichment methods. The genes in the clusters are also often tested for their enrichment with differentially expressed genes from transcriptomic analysis, as illustrated in Figure 2B. Based on these results, other non-differentially-expressed genes in the enriched clusters can be implicated in the disease using 'guilt-by-association' approaches. The newly implicated genes may have clinical importance as potential therapeutic targets and biomarkers.

Despite the aphorism "correlation is not causation", partial yet informative insights can be gleaned from co-expression networks, such as underlying regulatory framework mediating the co-expression patterns. New methods based on partial-correlations, Bayesian, and graphical Gaussian models (Werhli et al., 2006) take into account local connectivity when estimating edge strengths and a few methods work by combining prior-knowledge of expression patterns of TFs with co-expression information (Huynh-Thu et al., 2010; Rotival and Petretto, 2014). Gene-gene interaction network methods like ARACNe (Margolin et al., 2006) and Context Likelihood of Relatedness (CLR) (Faith et al., 2007) attempt to better capture these regulatory associations by accounting for connections within a shared neighborhood of genes in order to infer the strength of a link between two genes. Applying these approaches in complex conditions, like a gene being regulated by many regulators, becomes more challenging. Inferring the direct regulatory influence of transcription factors on target genes is central to interpreting the regulatory networks. Concerted efforts to support network-inference, such as the DREAM5 benchmark challenge (Marbach et al., 2012), have summarized different strategies that can be employed to infer regulatory networks. The accuracy of reconstruction approaches is often tested by comparing the predicted networks with high-confidence transcription factor binding data (He and Tan, 2016). However, integrating multi-omic data into these models to understand the pathobiology of disease states is an open challenge. Methods like CellNet (Cahan et al., 2014) an extension of CLR, and MOGRIFY (Rackham et al., 2016) take into account differentially expressed genes within the co-expression network framework in order to predict cellular reprogramming by transcription factors. Thus, co-expression methods have also been used to infer regulatory networks and to delineate the influence of regulatory genes, such as transcription factors, on their targets. However, obtaining condition-specific gene regulatory networks requires information regarding transcription factor binding activity in the given context. We will review methods that utilize TF binding information in the next section.

To summarize, inferring disease-specific information from GCN is possible from co-expressed or co-regulated clusters, differentially expressed and co-expressed genes, as well as the topological and functional properties of these. Biomedical big data measuring the transcriptome is highly leveraged by GCNs. For example, human tissue-specific GCNs have been constructed and analyzed (Pierson et al., 2015) using consortium data such as GTEx (Mele et al., 2015). These analyses revealed that genes with tissue-specific function are not hubs but connect to tissue-specific transcription factor hubs. Explorations using relative isoform ratios (RNA transcripts from same genes with different exons removed) and splicing data revealed distinct co-expression relationships unique to the tissues (Saha et al., 2017). Tissue specificity of gene co-expression networks have also been assessed in rats (Xiao et al., 2014), humans (Farahbod and Pavlidis, 2018; Kogelman et al., 2016; Ni et al., 2016; Prieto et al., 2008; Xiao et al., 2014), bats (Rodenas-Cuadrado et al., 2015), and plants (Aravind, 2000). Similarly, TCGA data has been analyzed using WGCNA in order to study the system-level properties of prognostic genes (Yang et al., 2014). Similar to gene co-expression, protein co-abundance networks can also be used to pinpoint influential proteins as potential regulators of the observed phenotype, and have been used to study inflammation (Halu et al., 2018), HCV infections (McDermott et al., 2012), and cancer, including breast cancer (Ryan et al., 2017) and glioblastoma (Kanonidis et al., 2016).

## 4.3 PARADIGM III: Inferring Phenotype specific Gene Regulatory Networks

In the previous sections, we studied various ways to construct networks and integrate molecular data to extract phenotype-specific biology in the form of gene prioritization, disease modules, or therapeutic targets. Those included immutable PPIs allowing disease-specific information to be embedded onto them and organic ways to model disease-specific information using co-expression networks. Here, separate networks are built for each phenotype which may be case-control, disease-specific, tissue or cell-specific, sex-specific, or for different disease subtypes. The network comparison model stems from the axiom of 'differential networking' over 'differential expression.' Many examples of differential networking can be found, including the INtegrated DiffErential Expression and Differential network analysis (INDEED) (Zuo et al., 2016) and DICER (Amar et al., 2013) algorithms. In this paradigm, we aim to discuss ways of leveraging phenotype-specific biomedical information to construct condition-specific gene regulatory networks. In principle, gene co-expression networks can also be phenotype-specific and can be used to infer condition-specific signals, but they lack the underlying set of canonical interactions unlike gene regulatory networks which include protein-DNA information in the form of TF binding information.

Instead of combining data from cases and controls to obtain key molecular elements, such as differentially-expressed genes or genes annotated to GWAS SNPs, in this paradigm the data is used to construct separate networks for each of the conditions. This construction of phenotype specific networks helps to mitigate systematic experimental biases and errors in both conditions (de la Fuente, 2010; Ideker and Krogan, 2012). It allows the comparison of networks to help uncover the specific rewiring of pathways, such as those induced by disease, pharmacological treatment (Bandyopadhyay et al., 2010), or environmental stimuli. GCNs can also be constructed in a phenotype-specific manner, as seen in the previous section. In Figure 2C, we depict an approach where phenotype-specific networks are constructed to uncover differentially targeted interactions. In this section, we focus on transcriptional regulatory networks that depend not only on co-expression, but also on modeling the binding propensities of TFs. These networks may also incorporate other multi-omic data to obtain condition-specific regulatory models.

The primary benefit of comparing phenotype-specific networks, particularly in GRNs, is to better delineate the role of genes in each condition. The 'rewiring' of the TFs targeting each of the genes can be tracked and the perturbations leading to these changes can convey information regarding the mechanistic underpinnings of the observed phenotype. An apt extension of 'differential networking' to the transcriptional regulatory network framework is 'differential targeting', which captures the highly dynamic nature of gene regulation. Changes in network topology, driven by underlying condition-specific data, can yield valuable insights and help to identify driver nodes and network biomarkers, such as a set of strengthened or weakened interactions between TF and target genes in the context of disease.

We review the Passing Attributes between Networks for Data Assimilation (PANDA) algorithm (Glass et al., 2013) as an exemplary method for constructing condition-specific regulatory networks, allowing for robust differential targeting analysis. PANDA is initiated by constructing a prior regulatory network consisting of potential routes for communication by mapping transcription factor motifs to a reference genome and assigning them to genes if they are in the regulatory region of the genes. PANDA then integrates other sources of information to iteratively optimize the flow of information through the network, modifying the prior to obtain a condition-specific regulatory network. The phenotype-specific regulatory networks are then compared to identify the structures most affected by this 'rewiring' and their biological significance. PANDA models the interactions between transcription factors based on the following principles. Firstly, if two transcription factors have a similar targeting profile, i.e., target similar genes or have binding motifs in the promoters of the same genes, they are more likely to physically interact or be members of the same TF complex (Guo et al., 2016; Hemberg and Kreiman, 2011). Cooperative binding of TFs is found to be evolutionarily constrained and conserved (Goke et al., 2011; He et al., 2011), and impacts crucial eukaryotic functions (He et al., 2011; Hochedlinger and Plath, 2009; Will and Helms, 2014; Wilson et al., 2010). Likewise, if two genes are targeted by the same set of TFs, these genes are likely to share similar expression patterns (Kim et al., 2006; Marco et al., 2009; Yu et al., 2003), or be part of the same functional module (Feldman et al., 2008; Goh et al., 2007). For this purpose, PANDA incorporates PPI networks to determine the 'responsibility' of TFs co-binding based on shared targets. It also uses GCNs to determine the 'availability' of genes to be simultaneously co-regulated, as evidenced by common co-expression. A vital component in PANDA is a 'prior' network composed of all potential regulatory routes based on the existence of binding sites for TFs in the regulatory regions of genes. All three ingredients (PPI, GCN, and a network prior) are then assimilated to uncover consistent patterns among these networks using a message-passing framework similar to affinity-propagation (Frey and Dueck, 2007). The outcome is a network elucidating the edges that form self-consistent modules, identifying relevant biological processes.

The phenotype-specific applications of PANDA are broad and include the comparison of disease and control networks in both complex diseases and cancers. For example, PANDA has been used to identify potential drug targets in ovarian cancer subtypes (Glass et al., 2015). Comparing PANDA networks between poor and good responders to asthma therapies identified potential transcriptional mediators of corticosteroid response in asthma (Qiu et al., 2018). The role of serotonin (5HT) dysregulation in mitral valve disease was explored using PANDA to find upregulation in 5HTR2B expression and increase 5HT receptor signaling (Driesbaugh et al., 2018). The effect of weight-loss on decreased risk of colorectal cancer was evaluated by applying PANDA to gene expression data on rectal mucosa biopsies (Vargas et al., 2016). In cancer research, PANDA network analysis in triple-negative breast cancer (TNBC) identified new core modules of functionally essential TFs and genes in cancer cells (Min et al., 2017). PANDA has also been used to investigate non-epithelial cancers like glioma to identify prognostic biomarkers mainly concerning mesenchymal signatures (Celiku et al., 2017). Sexual dimorphism, where the phenotypes are males and females, is another area where PANDA has been applied extensively, from sex-related targeting differences in COPD (Glass et al., 2014), colorectal cancer (Lopes-Ramos et al., 2018), and understanding crucial sex-related differences in various tissues in the human body (Chen et al., 2016). Differences between cell-lines and their host tissues have also been investigated using PANDA (Lopes-Ramos et al., 2017).

The issue of tissue-specificity can also be addressed by the paradigm of condition-specific networks, where the phenotype is the tissue or cell type. Various methods use gene expression data with regression trees (Huynh-Thu et al., 2010) or consider the context of pathways (Jambusaria et al., 2018). Enhancer and promoter data (Marbach et al., 2016) have been used to construct tissue-specific networks in humans and plants (Huang et al., 2018a). Using GTEx transcriptome data, PANDA has been used to construct gene regulatory networks for 38 distinct human tissues (Sonawane et al., 2017). This analysis assessed the inter-relationship between tissue-specific genes and TFs based on expression data and tissue-specific interactions and the topological positions of functionally important genes in respective tissues. This study also used network centrality measures like betweenness and degree to assess the topological properties of the nodes to identify rewiring around these genes in various tissues. Another significant contribution of this work is the elucidation of the tissue-specific regulatory roles of transcription factors, which were found to be independent of their expression levels. Instead, transcription factors appeared to mediate critical tissue-specific processes through subtle shifts in the gene regulatory networks, providing functional redundancy and, as a consequence, phenotypic stability of tissues.

## 5  Conclusion and future directions

Above we reviewed a limited set of network medicine philosophies that seek to integrate biomedical big data to uncover meaningful biology. Network medicine approaches provide customized and optimized ways to leverage biomedical data. The choice of these appropriate network method is largely dictated by the underlying biological inquiry, hypotheses, study design, and available data. Although this review is not meant to be exhaustive, our intent was to give a essence of how biomedical data requires a nuanced approach when selecting network analyses and provide a resource for both network scientists and biologists to better understand the lexicon of network modeling of biomedical data.

We believe that network medicine approaches will be vital in the future with the increasing emergence of diverse technologies, multi-omic data types, deeper levels of inquiry from tissues to cellular levels, platforms that include large amounts of publicly available biomedical data, and efforts in precision medicine, which aim to find the right drugs for the right patients at the right time. There is a growing realization that genomics is only a part of the story when it comes to cancer and other complex diseases. The field is working to augment genetic information

(mutations, deletions, and other somatic genetic alterations) with other omics data, such as epigenomics (methylation, non-coding RNAs, histone modifications, chromatin structures), proteomics (in vitro studies on proteins), and lipidomics (survey of cellular lipids), to name a few. The network medicine framework presents a promising way of thinking about and integrating these heterogeneous data types by elucidating their mutual influences to help explain disease etiologies and cellular functions and providing the basis for personalized therapeutics.

Multi-omics data integration using networks has already started gaining a wide amount of attention in the scientific community (Gligorijevic and Przulj, 2015; Hasin et al., 2017; Huang et al., 2017; Tuncbag et al., 2016; Yugi et al., 2016). Moreover, relatively newer network tools like multiplex networks (Didier et al., 2018), network fusion (Wang et al., 2014), more innovative community detection strategies (Gligorijevic et al., 2016), higher order structural modularity (Didier et al., 2018), have the potential to be applied to these problems to gain an even deeper and more nuance understanding of biological systems. Multilayer network approaches (De Domenico et al., 2015) for human diseases have unraveled important associations between rare and complex diseases (Halu et al., 2017). Despite several open challenges (Stegle et al., 2015; Ziegenhain et al., 2017), new technologies like single-cell transcriptomics (Hon et al., 2018), have started to be used to construct GRNs (Fiers et al., 2018; Herbach et al., 2017) and cell-specific coactivation networks (Ghazanfar et al., 2016). As the field of network medicine moves forward, one thing that is required more than ever before is the development of methods for systematically validating network predictions. Such validation will provide a greater confidence in network predictions and facilitate their incorporation into translational medicine. We also think active trans-disciplinary collaboration between biologists and scientists from the field of complex networks is required to infuse the field of network medicine with novel algorithms and innovative strategies. The application of network methods to biomedical data presents a great opportunity to test and improve upon the tools originating from the general field of complex networks. We also take this opportunity to thank the many experimental biologists whose operose efforts have led to the generation of the vast amount of invaluable biomedical data, and to the numerous individuals who have donated their data for the sake of science.

## 6 Conflict of Interest

*The authors declare that the research was conducted in the absence of any commercial or financial relationships that could be construed as a potential conflict of interest.*

## 7 Author Contributions

All authors listed contributed substantially to the intellectual, writing and editing, and approved the manuscript for publication.

## 8 Funding

K. Glass is supported by the NIH/NHLBI through K25HL133599. We acknowledge the support by National Institutes of Health (NIH) grants R01 HL118455-04-1 and P01 HL13285. The funders had no role in study design, data collection and analysis, decision to publish, or preparation of the manuscript.

## 9 Acknowledgments

ARS would like to thank John Quackenbush for inspiration of the 3 paradigms discussed above with Trevor Leonardo and Rebecca Burkholz for critical reading of the manuscript. We also thank members of the Quackenbush and Sharma labs for many fruitful discussions.

Figure legends:

Figure 1: Overview of network medicine approach depicting various biomedical data types discussed at length in the manuscript, along with network representations that simplify different components of multiple omics data from the genome, transcriptome, proteome, and metabolome as nodes that are connected by links (edges). Combining biomedical data with the appropriate network modeling approach allows derivation of disease associated information and outcomes like biomarkers, therapeutics targets, phenotype-specific genes and interactions, and disease subtypes.

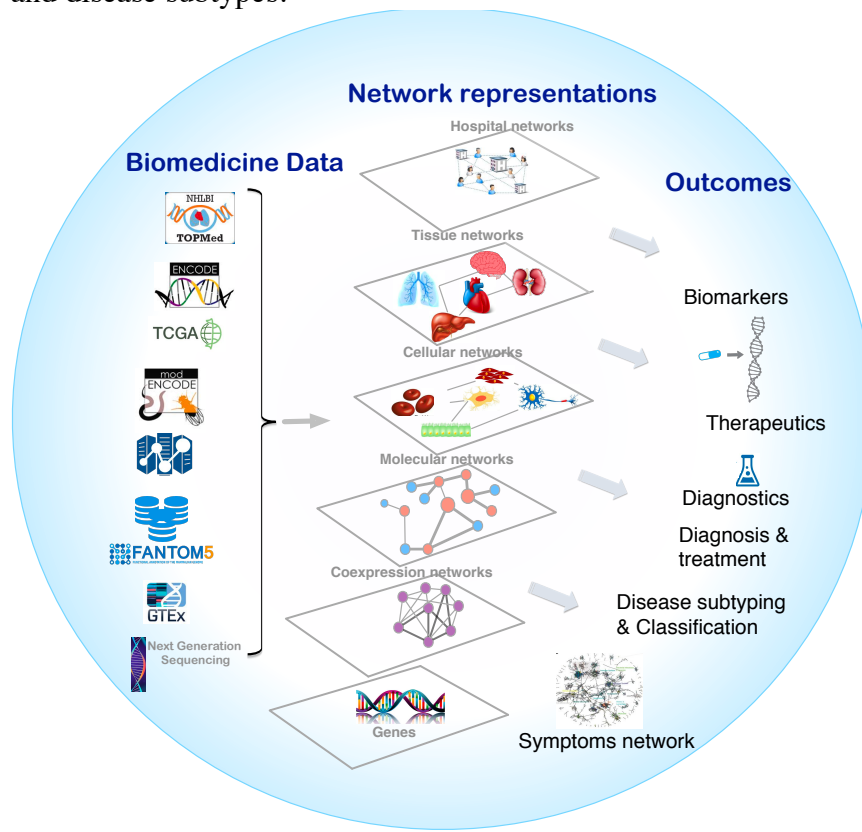

Figure 2: Schematic of three paradigms for combining biological networks with phenotype-specific biomedical data, such as a set of disease genes and transcriptomic profiles for case and control groups. (a) Identification of disease associated network components within the interactome, (b) Co-expression based network modeling to identify disease biomarkers, (c) Constructing phenotype-specific GRNs to identify perturbations and condition-specific regulatory changes.

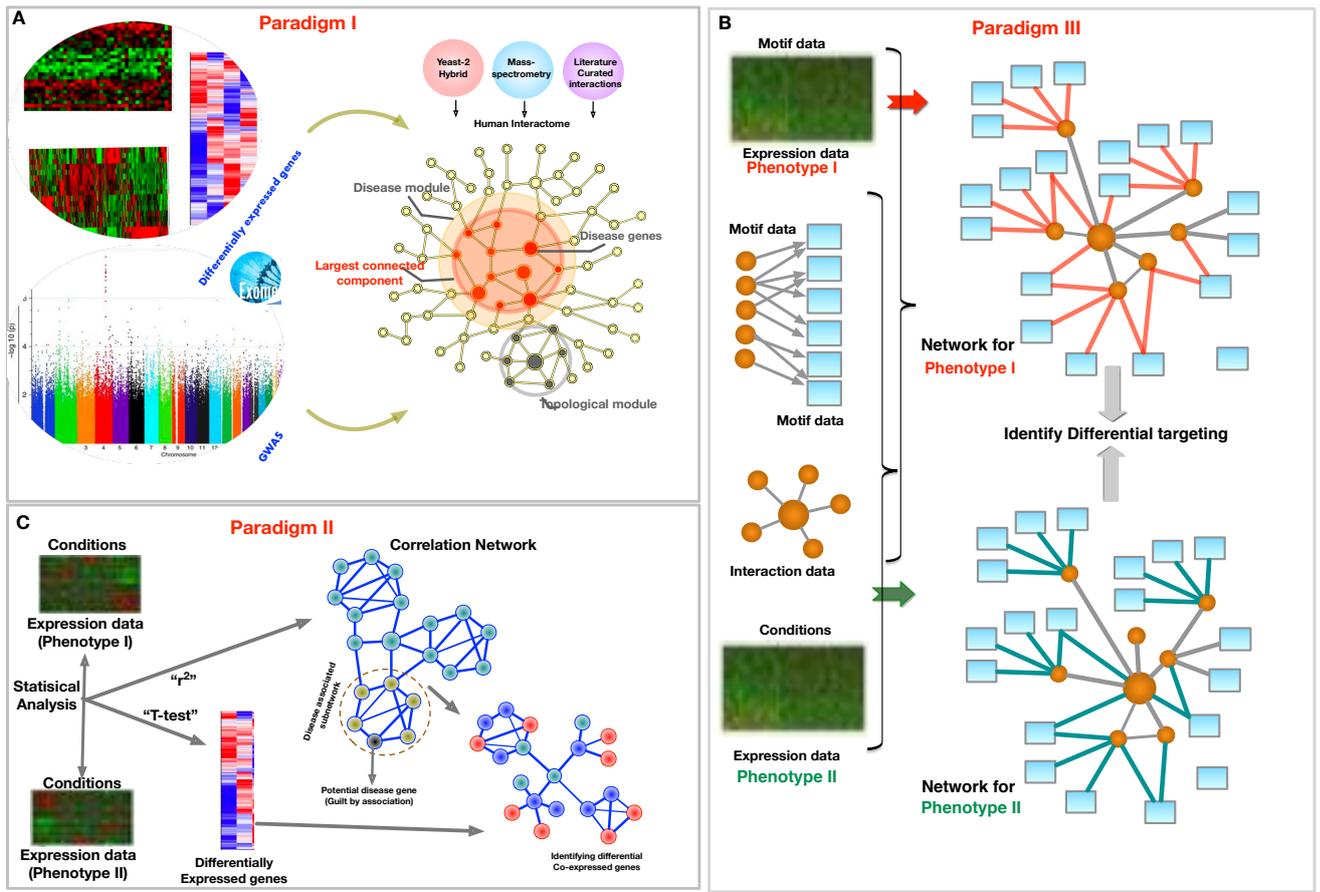

## List of important abbreviations

| PPI | Protein-Protein Interactions |
|---|---|
| NGS | Next Generation Sequencing |
| CNV | Copy Number Variation |
| SNP | single nucleotide polymorphism |
| TCGA | The Cancer Genome Atlas |
| ENCODE | ENCyclopedia Of DNA Elements |
| (modENCODE) | model organism ENCyclopedia Of DNA Elements |
| FANTOM5 | Functional ANnoTation Of Mammalian genome |
| GTEx | Genotype-Tissue Expression |
| TOPMed | Trans-Omics for Precision Medicine |
| HPA | Human Protein Atlas |
| HCA | Human Cell Atlas |
| HMP | Human Microbiome Project |
| GCN | Gene Co-expression Networks |
| GRN | Gene Regulatory Networks |